\documentclass[aps,twocolumn,showpacs,superscriptaddress,amssymb]{revtex4-1}
\date{\today}
\usepackage{graphicx,threeparttable,amssymb}
\usepackage{dcolumn}
\usepackage{bm}
\usepackage{color}

\begin{document}

\title{Uncertainty decomposition method and its application in the liquid drop model}

\author{Cenxi Yuan}
\email{yuancx@mail.sysu.edu.cn} \affiliation{Sino-French Institute
of Nuclear Engineering and Technology, Sun Yat-Sen University,
Zhuhai, 519082, Guangdong, China}

\begin{abstract}
A method is suggested to decompose the statistical and systematic uncertainties from the residues between the calculation of a theoretical model and the observed data. The residues and the parameters of the model can be obtained through the standard statistical fitting procedures. The present work concentrates on the decomposition of the total uncertainty, of which the distribution corresponds to that of the residues. The distribution of the total uncertainty is considered as two normal distributions, statistical and systematic uncertainties. The standard deviation of the statistical part, $\sigma_{stat}$, is estimated through random parameters distributed around their best fitted values. The two normal distributions are obtained by minimizing the moments of the distribution of the residues with the fixed $\sigma_{stat}$. The method is applied to the liquid drop model (LD). The statistical and systematic uncertainties are decomposed from the residues of the nuclear binding energies with and without the consideration of the shell effect in LD. The estimated distributions of the statistical and systematic uncertainties can well describe that of the residues.

The normal assumption of the distribution of the statistical and systematic uncertainties is examined through various approaches. The comparison between the distributions of the specific nuclei and those of the statistical and systematic uncertainties are consistent with the physical considerations, although the latter two can be obtained without the knowledge of these considerations. Such as, the LD are more suitable to describe the heavy nuclei. The light and heavy nuclei are indeed distributed mostly inside the distributions of the statistical and systematic uncertainties, respectively. The similar situation are also found for the nuclei close to and far from shell. The present method is also performed to nuclei around stability line. The results are used to investigate all measured nuclei, which show the usefulness of the UDM in the exploration of the unmeasured nuclei.
\end{abstract}


\maketitle

\section{\label{sec:level1}Introduction}
The estimation of the uncertainty of a theoretical model is of great importance to
evaluate the predicted ability of the model~\cite{doba2014JPG}. The standard statistical methods, such as the least square and $\chi^{2}$ fitting, are widely used in the parametrization of various models. Specially in nuclear physics, the methods are used to control the validity of the data fitting procedure in the liquid drop model (LD)~\cite{myers1966}, finite-range droplet model~\cite{moller1995}, Lublin-Strasbourg Drop model~\cite{pomorski2003}, Woods-Saxon model~\cite{dudek1978,dudek1979}, and Skyrme (like) force~\cite{bartel1982,chabanat1997,chabanat1998}. Various of observed data can be considered in the fitting procedure, such as the nuclear mass, radius, single particle energies, deformations, and so on. Taking nuclear mass models for example, the total uncertainties are obtained to be around $0.5$ MeV from the fitting procedure in the finite-range droplet model~\cite{moller1995}, the Lublin-Strasbourg Drop model~\cite{pomorski2003}, the Hartree-Fock-Bogoliubov model~\cite{Goriely2009,Goriely2013}, and the Weizs\"{a}cker-Skyrme mass model~\cite{liu2011,Wang2014}.
The recent version of Weizs\"{a}cker-Skyrme mass model considers the effect of the surface diffuseness, which is important for nuclei with extreme isospin~\cite{Wang2014}. Such effect is especially evident for the extremely neutron-rich nuclei, because the valence neutron may extend very far due to the lack of the Coulomb barrier. Recent investigation on the heaviest known neutron-halo nuclei, $^{22}$C, show that the upper limit of the radius is a key characteristic of the two-neutron halo~\cite{Suzuki2016}.

It is of great interesting to investigate the details of the total uncertainty obtained from fitting procedures.
The total uncertainty normally comes from three parts, the model, the experiment, and the numerical method~\cite{doba2014JPG}.
The uncertainty from the model consists two parts,
the statistical uncertainty from the not exactly determined parameters and the systematic uncertainty from the deficiency of the model.
The systematic uncertainty is hard to be estimated because its origin is the deficiency of the theoretical model~\cite{doba2014JPG}.
In Ref.~\cite{erler2011}, the systematic uncertainty are obtained by comparing a variety of models.
Two illustrative examples are given to estimate the systematic uncertainty by analysing the residues~\cite{doba2014JPG}.

Here we suggest a practical method to decompose the statistical and systematic uncertainties from the total uncertainty and its distribution in one model in the case of large sample. In the case of large sample, both distributions of statistical and systematic uncertainties are considered as normal distributions. The moments of the residues are used to constrain the normal function. In the parametrization of the model, the uncertainty of each model parameters is obtained through the standard fitting procedure. The standard deviation of the statistical uncertainty is estimated through the randomly generated parameters following the normal distribution defined by their uncertainties.

To decompose the total uncertainty, as an example, a possible choice is applying to the LD because of its simplicity. The uncertainty of the model parameters can be easily obtained through the linear fitting procedure. One of the well known deficiencies of the LD is the lack of the shell effect. It is helpful for further discussion when the present decomposition method applying to the LD with and without the shell effect.

\section{\label{sec:level2}Theoretical Framework}

An observed data $Y$ is described by a model with a few parameters $(X_{1},X_{2}...)$. After fitting procedure, the values and uncertainties of all parameters are obtained, $X_{i}$ and $\sigma_{i}$ for the $i$th parameters. Its total uncertainty, defined by the residue $e=Y(X_{1},X_{2}...)-Y(Expt.)$, includes three parts, the uncertainties from the model, the experiment, and the numerical method~\cite{doba2014JPG}. In the present study, only uncertainties from the model are considered for simplicity, including statistical and systematic uncertainties. It is reasonable for the LD because the experimental uncertainty of the binding energy is generally very small~\cite{audi2012} and the numerical uncertainty of a linear and analytical model is negligible.

The distribution of the total uncertainty is the sum of the distributions of the statistical and systematic uncertainties.
In the case of a large sample, it is reasonable to suppose that the statistical and systematic uncertainties follow the normal distribution, although not exactly. A normal distribution is labeled as $N(m,\sigma)$, with the mean value $m$ and the standard deviation $\sigma$.
The distribution of total uncertainty is:
\begin{eqnarray}\label{residue}
    f(e) & =& f(stat)+f(syst) \nonumber \\
    & =& \frac{1}{2}N(m_{stat},\sigma_{stat})+\frac{1}{2}N(m_{syst},\sigma_{syst}).
\end{eqnarray}
where $\frac{1}{2}$ is the normalized factor. The mean values $m_{stat}$ and $m_{syst}$ are generally separated, which is rarely discussed in the previous works.

The moments are important quantities in the description of a distribution, such as the mean value (first moment), variance (second moment), skewness (third moment), kurtosis (fourth moment). Applying the calculations of moments to Eq.~(\ref{residue}):
\begin{eqnarray}\label{residue3}
    m(e) & =& \frac{1}{2}(m_{stat}+m_{syst}) \nonumber \\
    \sigma^{2}(e)&=& \frac{1}{2}(m_{stat}^{2}+\sigma_{stat}^{2}+m_{syst}^{2}+\sigma_{syst}^{2}) \nonumber \\
    p_{3}(e) &=& \frac{1}{2}(m_{stat}^{3}+3m_{stat}\sigma_{stat}^{2}+m_{syst}^{3}+3m_{syst}\sigma_{syst}^{2})\nonumber \\
    p_{4}(e) &=& \frac{1}{2}(m_{stat}^{4}+6m_{stat}^{2}\sigma_{stat}^{2}+3\sigma_{stat}^{4}+ \nonumber \\
    && m_{syst}^{4}+6m_{syst}^{2}\sigma_{syst}^{2}+3\sigma_{syst}^{4})
\end{eqnarray}
The moments in the left hand side are calculated through the distribution of $Y(X_{1},X_{2}...)-Y(Expt.)$.
The right hand side is obtained through the properties of the normal distribution.
In principal the mean values and variances of the statistical and systematic uncertainties can be obtained  through Eq.~(\ref{residue3}).
However it sometimes has no physical solution because the normal distribution assumption is not exactly.

The variance of the statistical uncertainty can be simulated through $X_{i}$ and $\sigma_{i}$:
\begin{eqnarray}\label{sigmastat}
    \sigma_{stat}^{2}&=& \frac{\Sigma_{k=1}^{M}[Y(X_{i}^{\prime})_{k}-Y(X_{i})_{k}]^{2}}{M}.
\end{eqnarray}
$X_{i}^{\prime}$ is randomly generated through a normal distribution $N(X_{i},\sigma_{i})$. The statistical uncertainty comes from the uncertainty of the parameters of the model. For the $k$th term, we randomly select one $Y_{k}$ and one parameter $X_{i}$ from all possible candidates, while other parameters the same as the best fitted values. The number $M$ is chosen to be sufficient large comparing with the number of the observed data. Such procedure simulates the deviation comes from the uncertainty of the parameters of the model. It is the estimation of the variance of the statistical uncertainty.

Together with the first two equations in Eq.~(\ref{residue3}), one can express $m_{syst}$ and $\sigma_{syst}$ by $m(e)$, $\sigma(e)$, $\sigma_{stat}$, and $m_{stat}$. Only one unknown remains in the latter two equations in Eq.~(\ref{residue3}). One can calculate the $p_{3}$ and $p_{4}$ as the function of $m_{stat}$ and minimize,
\begin{eqnarray}\label{criteria}
   \triangle p&=&|p_{3}^{1/3}(e)-p_{3}^{1/3}(m_{stat})|+|p_{4}^{1/4}(e)-p_{4}^{1/4}(m_{stat})|, \nonumber
\end{eqnarray}
to estimate the value of $m_{stat}$, which is the criteria for the present study. The method discussed is labeled as the uncertainty decomposition method (UDM).

In the present work, the UDM is applied to the LD. The LD is an empirical model describing the binding energies and other bulk properties of nuclei.
Microscopic approaches describe the binding energies of most nuclei with good accuracy, such as the energy density functional theory~\cite{erler2011} and the Hartree-Fock-Bogoliubov method~\cite{Goriely2009}. Some other microscopic approaches, such as the nuclear shell model, concentrate on the light and medium mass nuclei. Our previous works show that the shell model can give a precise description on the light nuclei from stability line to both the neutron and proton drip line~\cite{yuan2012}. The LD can give well description on the binding energies of nuclei~\cite{moller1995,moretto2012} with the standard shell correction procedure~\cite{stru1967}.

The original LD mass formula includes the volume energy, the surface energy, the Coulomb energy, the volume term of proton-neutron asymmetry energy,
and the paring energy~\cite{heydebasic}. Many additional terms are introduced to include more physical effect.
Such as, the surface energy of proton-neutron asymmetry is introduced to the LD mass formula~\cite{myers1966}.
The LD mass formula is given as:
\begin{eqnarray}\label{LD6}
 BE(A,Z)_{LD6} &=& a_{v}A-a_{s}A^{2/3}-a_{c}Z(Z-1)A^{-1/3} \nonumber \\
    & & -a_{a}^{v}I(I+1)/A+a_{a}^{s}I(I+1)/A^{4/3} \nonumber \\
    & & +\delta a_{p}A^{-1/2},
\end{eqnarray}
where $I=|A-2Z|$ and $\delta=1,0,-1$ for even-even, odd-even, and odd-odd nuclei, respectively.
From and after we label Eq.~(\ref{LD6}) as LD6 because of its six parameters. It should be noted that there are several forms of the surface asymmetry term when introducing to the LD~\cite{jiang2012}.
These six parameters are considered to be the most important terms for nuclear binding in a macroscopic view and reproduce experimental binding energies,
generally speaking, within the precision of $1\%$. The largest deviation comes from the lack of the shell effect.

Strutinsky procedure is the standard method to introduce the shell effect to the LD~\cite{stru1967}.
A shell correction term is presented as the function of the number of the valence nucleons respected to the closed shell~\cite{Dieperink2007}:
\begin{eqnarray}\label{shell1}
 BE(N_{p},N_{n})_{shell} &=& a_{1}F_{max} + a_{2}F^{2}_{max},
\end{eqnarray}
where $F_{max}=(N_{p}+N_{n})/2$ with $N_{n(p)}$ is the number of the valence neutron (proton) respected to the nearest closed shell.
It is obvious that in an illustrative view the shell correction is related to the number of the valence nucleons.
Here, we use an illustrative function for shell correction to ignore the numerical uncertainty and for further discussion:
\begin{eqnarray}\label{shell21}
    G(Z,N)&=&(Z+N)\Sigma_{Z_{0},N_{0}}e^{-\frac{(Z-Z_{0})^{2}+(N-N_{0})^{2}}{a_{0}^{2}}} \nonumber
\end{eqnarray}
where $Z_{0}$ and $N_{0}$ are the spin-orbit magic numbers, $28$, $50$, $82$ for both proton and neutron, and $126$ for neutron, and
\begin{eqnarray}\label{shell2}
    BE(Z,N)_{shell}&=& a_{sh}G(Z,N),
\end{eqnarray}
with $a_{shell}$ the strength of the shell correction. The function considers the shell effect in an illustrative way that the nuclei around doubly magic nuclei get extra binding energy. The shell correction decreases when nuclei go far away from the doubly magic nuclei,
with $a_{0}^{2}$ the scale of the distance. The exclusion of the magic numbers $8$ and $20$ is because the LD6 do not show the systematic necessity of
the extra binding energy for the nuclei around these two magic numbers. The possible reason is that the LD works insufficiently in the light region.

With the shell correction term Eq.~(\ref{shell2}), the LD mass formula can be written as:
\begin{eqnarray}\label{LD8}
BE(A,Z)_{LD8} &=& a_{v}A-a_{s}A^{2/3}-a_{c}Z(Z-1)A^{-1/3} \nonumber \\
    & & -a_{a}^{v}I(I+1)/A+a_{a}^{s}I(I+1)/A^{4/3} \nonumber \\
    & & +BE(Z,A-Z)_{shell}+\delta a_{p}A^{-1/2},
\end{eqnarray}
which is labeled as LD8 in the later discussion. In the following section, the UDM is applied to the LD.

\section{\label{sec:level3}Uncertainty Decomposition of Liquid Drop Model}

\begin{table*}
\caption{\label{para} Six sets of parameters including standard deviations of each parameter in parentheses for the LD6 and LD8 fitted to AME1995, AME2003, and AME2012, the mean value $m(e)$, standard deviations $\sigma(e)$, skewness to the power of $1/3$, $p_{3}^{1/3}(e)$, and kurtosis to the power of $1/4$, $p_{4}(e)$, of the residues of the binding energies. All data are in the unit of MeV except the dimensionless quantity $a_{0}^{2}$.}
\begin{ruledtabular}
\begin{tabular}{cccccccccccccc}
 Model &data  &$a_{v}$  & $a_{s}$  & $a_{c}$  & $a_{av}$ & $a_{as}$  & $a_{p}$ & $a_{sh}$ & $a_{0}^{2}$ & $m(e)$ & $\sigma(e)$  &$p_{3}^{1/3}(e)$ &$p_{4}^{1/4}(e)$\\
\hline
 LD6 &AME1995& 15.705(25)    & 17.933(80)    &   0.7060(18)  &   29.21(21)&   38.90(1.11)& 11.65(86) &          &     & 0.01& 2.61 & -2.74 & 3.80  \\
 LD6 &AME2003& 15.683(23)    & 17.851(73)    &   0.7052(16)  &   28.89(18)&   37.47(95)  & 12.44(76) &          &     & 0.08& 2.47 & -2.62 & 3.69  \\
 LD6 &AME2012& 15.671(21)    & 17.807(69)    &   0.7041(14)  &   28.99(16)&   37.73(78)  & 12.14(74) &          &     & 0.08& 2.52 & -2.71 & 3.78  \\
 LD8 &AME1995& 15.674(14)    & 17.923(44)    &   0.7012(10)  &   29.70(12)&   40.84(62)  & 11.90(48) &0.0615(10)&40   &-0.04& 1.44 & -1.02  & 1.86  \\
 LD8 &AME2003& 15.668(13)    & 17.894(41)    &   0.7012(9)   &   29.52(10)&   39.99(54)  & 12.33(43) &0.0624(9) &40   & 0.04& 1.39 & -0.95  & 1.82  \\
 LD8 &AME2012& 15.656(12)    & 17.859(39)    &   0.7002(8)   &   29.53(9) &   40.01(43)  & 12.24(41) &0.0642(9) &40   &-0.03& 1.41 & -1.07  & 1.85  \\
\end{tabular}
\end{ruledtabular}
\end{table*}

\begin{figure}
\includegraphics[scale=0.25]{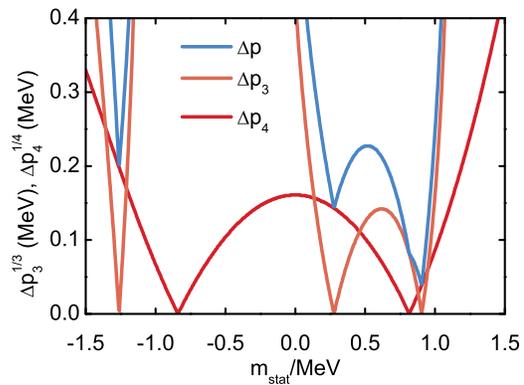}
\caption{\label{m1p3p4} (Color online) $\triangle p=|p_{3}^{1/3}(e)-p_{3}^{1/3}(m_{stat})|+|p_{4}^{1/4}(e)-p_{4}^{1/4}(m_{stat})|$, $\triangle p_{3}=|p_{3}^{1/3}(e)-p_{3}^{1/3}(m_{stat})|$, and $\triangle p_{4}=|p_{4}^{1/4}(e)-p_{4}^{1/4}(m_{stat})|$ as the function of $m_{stat}$, taking AME2012LD8 as an example.}
\end{figure}

\begin{table*}
\caption{\label{msd} Four estimated normal parameters, and the variance of the estimated uncertainties with six sets of the parameters obtained from the LD6 and LD8 fitted to AME1995, AME2003, and AME2012. All data are in the unit of MeV, except the $\frac{1}{2}(m_{stat}^{2}+\sigma_{stat}^{2})$ and $\frac{1}{2}(m_{syst}^{2}+\sigma_{syst}^{2})$ in MeV$^{2}$.}
\begin{ruledtabular}
\begin{tabular}{cccccccc}
 Model &data  &$m_{stat}$  & $\sigma_{stat}$  &$m_{syst}$  & $\sigma_{syst}$ & $\frac{1}{2}(m_{stat}^{2}+\sigma_{stat}^{2})$ &$\frac{1}{2}(m_{syst}^{2}+\sigma_{syst}^{2})$ \\
\hline
 LD6 &AME1995& 1.02    & 1.63    &   -1.00  &   2.98 &  1.84    &  4.95   \\
 LD6 &AME2003& 1.01    & 1.66    &   -0.86  &   2.77 &  1.89    &  4.22   \\
 LD6 &AME2012& 1.05    & 1.66    &   -0.88  &   2.85 &  1.93    &  4.45   \\
 LD8 &AME1995& 0.94    & 0.91    &   -1.03  &   1.17 &  0.86    &  1.22   \\
 LD8 &AME2003& 0.81    & 0.92    &   -0.74  &   1.33 &  0.75    &  1.16   \\
 LD8 &AME2012& 0.91    & 0.86    &   -0.96  &   1.22 &  0.78    &  1.20   \\
\end{tabular}
\end{ruledtabular}
\end{table*}

\begin{figure}
\includegraphics[scale=0.25]{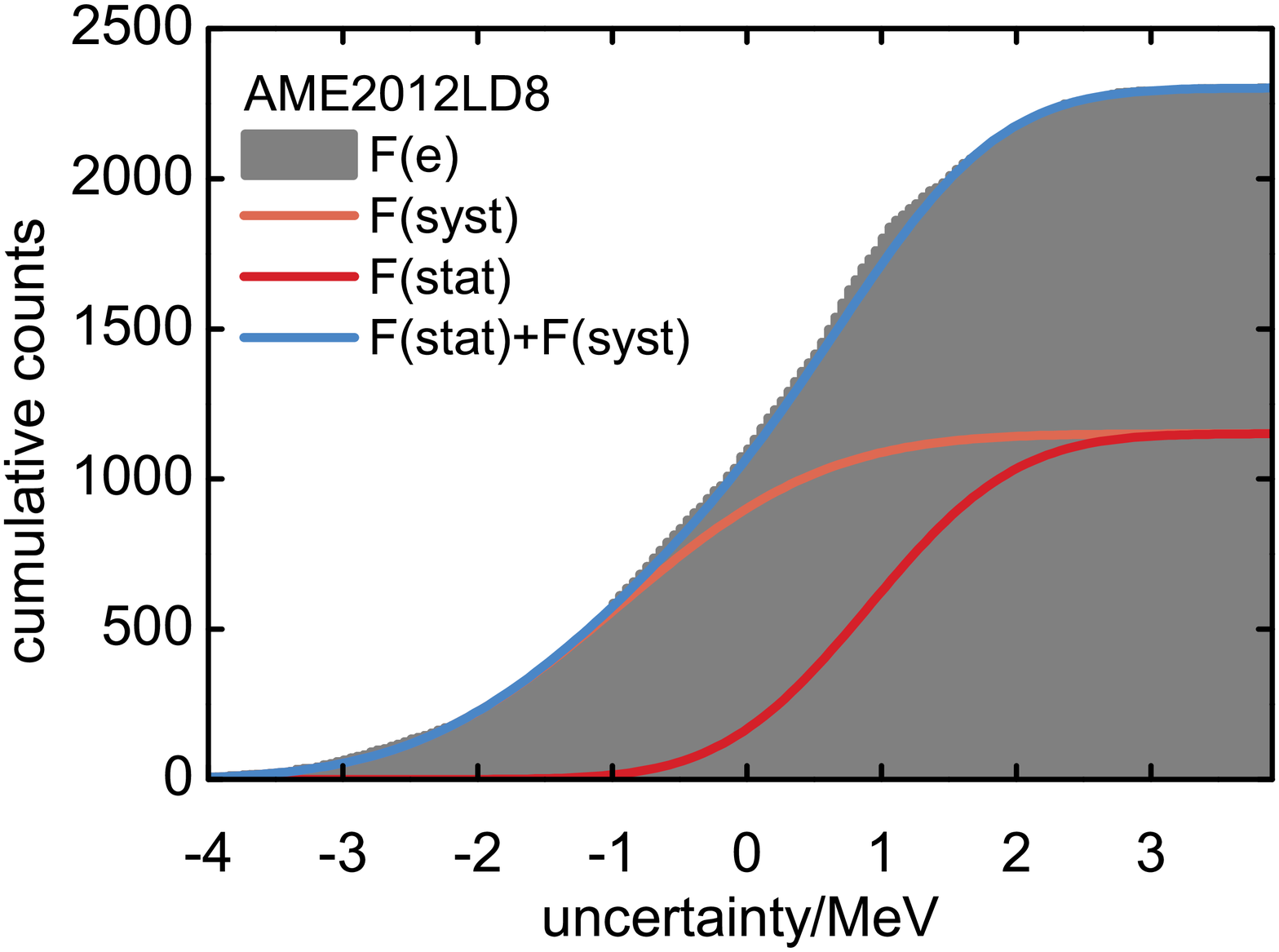}
\includegraphics[scale=0.25]{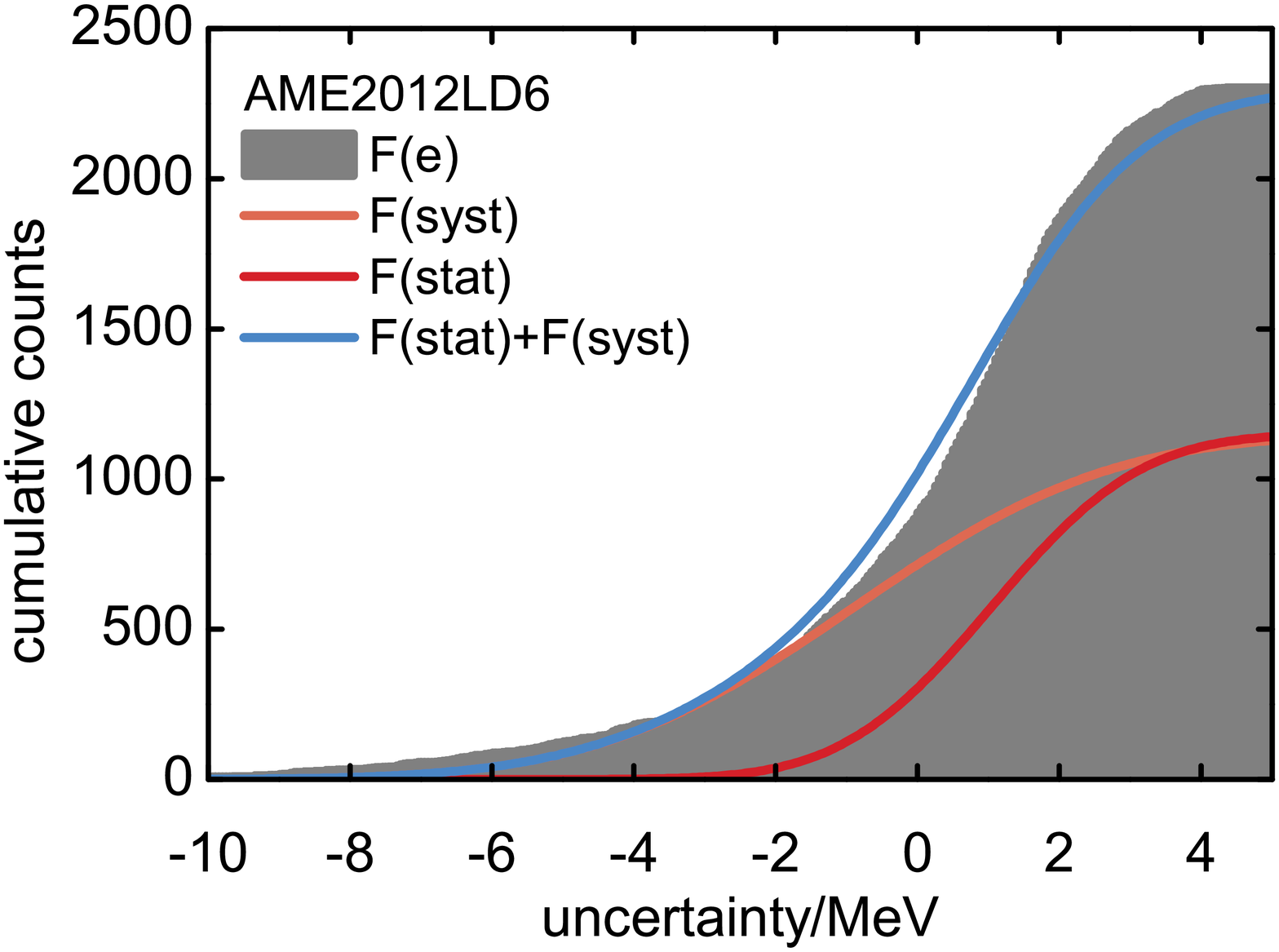}
\caption{\label{LDuncertainty} (Color online) The cumulative counts of the residues of the binding energies and the corresponding estimated uncertainties in AME2012LD6 and AME2012LD8. }
\end{figure}

We first determine the parameters of the LD6 and LD8 by fitting to three sets of the experimental data, AME1995, AME2003, and AME2012.
In the fitting, we only consider nuclei with experimental uncertainty smaller than $0.2$ MeV. Because the standard deviations of LD6 and LD8 are much larger than $0.2$ MeV, the experimental uncertainty is ignored in the following discussion. The light nuclei ($Z<7$ and $N<7$) are also excluded. Table~\ref{para} presents the parameters and their standard deviations obtained by the linear fitting. The standard deviations of parameter $a_{0}^{2}$, scaling the distance to doubly magic nuclei, is not given because it is fixed in the linear fitting. Actually, part of its uncertainty is included in the uncertainty of the strength of the shell correction, $V_{Z_{0},N_{0}}$. If we change the $a_{0}^{2}$ from $40$ to $30$ or $50$ and refit, the total uncertainty changes little ($<0.05$ MeV). All parameters keep almost the same (around $0.1\%$ change) except the $V_{Z_{0},N_{0}}$ (around $8\%$ change). In the following discussion, only the uncertainty of the first seven parameters in LD8 is investigated, with the fixed $a_{0}^{2}$. Three sets of the parameters and moments in LD6 or LD8 are close to each other indicating the LD gives the similar description for both early and newly discovered nuclei.

After the parameters and their uncertainties are obtained, the UDM introduced in the section~\ref{sec:level2} is used to decompose the statistical and the systematic uncertainties. Taking AME2012LD8 as an example, FIG.~\ref{m1p3p4} presents $\Delta p$, $\Delta p_{3}$ and $\Delta p_{4}$ as the functions of $m_{stat}$. There are three solutions for $\Delta p_{3}=0$ and two solutions for $\Delta p_{4}=0$. The solution $m_{stat}=0.91$ corresponds to the minimums of $\Delta p$. Thus we estimate the $m_{stat}$ to be $0.91$ and calculate corresponding $m_{syst}$ and $\sigma_{syst}$. The results are presented in the Table~\ref{msd}. The $\frac{1}{2}(m_{syst}^{2}+\sigma_{syst}^{2})$ and $\frac{1}{2}(m_{syst}^{2}+\sigma_{syst}^{2})$ are the variances of the statistical and systematic uncertainties, respectively. From the data AME1995 to AME2012, both the statistical and systematic uncertainties in LD6 or LD8 are closed to each other, which are around $1.9$ and $4.5$ MeV$^{2}$ in LD6, $0.8$ and $1.2$ MeV$^{2}$ in LD8. In Ref.~\cite{doba2014JPG}, it is explained that the statistical uncertainty has the same magnitude as the total uncertainty and is proportional to $\frac{N_{d}}{N_{d}-N_{p}}$, where $N_{d}$ and $N_{p}$ are the numbers of the data and parameters, respectively. The numbers of data selected from AME1995, AME2003, and AME2012 in present work are 1704, 2057, and 2302, respectively, which are much larger than the number of the parameters. The similar statistical and systematic uncertainties obtained from the different data sets are reasonable in the present fixed model with large samples. Both the statistical and systematic uncertainties reduce from the LD6 to LD8. The reduction of the statistical uncertainty is because of the increment of the number of the parameters. After an illustrative inclusion of the shell effect, the systematic uncertainty also reduces.

It is necessary to see whether the statistical and systematic function obtained from the UDM can describe the distribution of the residues. Figure~\ref{LDuncertainty} presents the cumulative counts of the distribution of the total and estimated uncertainties. It is seen that the estimated distribution from the LD8 give better description than that from the LD6. It indicates that the normal assumption of the systematic uncertainty is not exactly in the LD6 because of the lack of the shell effect. The binding energies of the doubly magic and nearby nuclei are generally much larger than the LD6 calculations, resulting the deviation from a normal distribution. From the LD6 results in FIG.~\ref{LDuncertainty}, around $5\%$ of the data have deviations smaller than $-5$ MeV (smallest one $-11.515$ MeV), but no deviations larger than $5$ MeV (largest one $4.388$ MeV). It is shown that the UDM can give an approximately solution for models with a large deficiency. Although the LD8 has certain deficiencies, such as the not exactly treatment of shell effect, the lack of the deformation and so on, the total uncertainty of the LD8 can be well described by the UDM. It means the normal assumption of the systematic uncertainty is reasonable which agrees with the central limit theorem. If the impact of each deficiency is small, the distribution of systematic uncertainty can be assumed to be normal distribution in the case of large sample. More discussions on the validity of the method are in the next section. The UDM gives nice description for the models with certain deficiencies. If the contribution of all deficiencies are smaller, characterized by the less variance and more symmetric distribution of the total uncertainty, the UDM works better as shown from the LD6 to LD8.

\section{\label{sec:level4}Validity of the Method}

\begin{figure}
\includegraphics[scale=0.30]{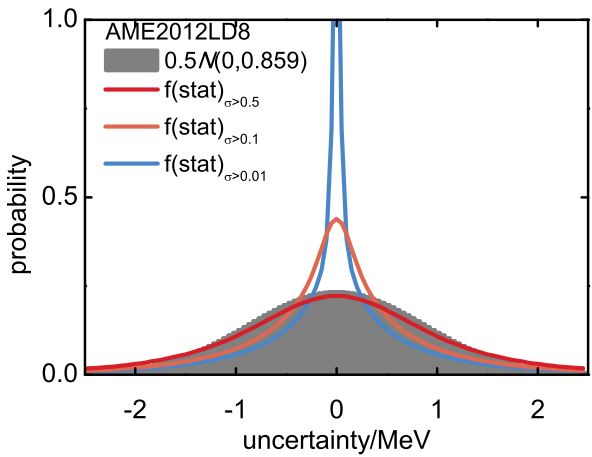}
\caption{\label{normaltest} (Color online) The estimated normal statistical distribution and the simulated distribution of Eq.~(\ref{LD8stat}) from AME2012LD8, when the latter ignoring terms with $\sigma=\sigma_{a_{i}}B_{i,Z,A}<0.01$, $0.1$, and $0.5$ MeV.}
\end{figure}

The UDM assumes that both the statistical and systematic uncertainties follow the normal distribution. The assumption should be tested to see its validity. The simulated distribution of the statistical uncertainty described in Eq.~(\ref{sigmastat}) can be exactly presented. If only the parameter $a_{v}$ changes, the simulated statistical residue is $(a_{v}-a_{v}^{\prime})A$ for certain nucleus ($A$, $Z$), where $a_{v}^{\prime}$ follows the normal distribution $N(a_{v}, \sigma_{a_{v}})$. The distribution of $(a_{v}-a_{v}^{\prime})A$ is a normal distribution $N(0, \sigma_{a_{v}}A)$. The simulated distribution of the statistical uncertainty is actually the summation of the normal distributions for all possible nuclei ($A$, $Z$) and parameters $a_{i}$:
\begin{eqnarray}\label{LD8stat}
f(stat) &=& \frac{1}{2}\frac{1}{N_{p}N_{d}}\Sigma_{Z,A}N(0, \sigma_{a_{i}}B_{i,Z,A}),
\end{eqnarray}
where $a_{i}$ and $B_{i,Z,A}$ are the parameter of the $i$th term in the LD and the corresponding function of $A$ and $Z$, respectively. Such as $B_{i,Z,A}$ is $A$ for volume term. $\frac{1}{2}\frac{1}{N_{p}N_{d}}$ is the normalized factor, where $N_{p}$ and $N_{d}$ are the numbers of the data and the parameters, respectively.

A normal distribution is used as the statistical distribution in the discussion of the section~\ref{sec:level3}. Figure~\ref{normaltest} presents the comparison between the simulated distribution and the normal statistical distribution obtained for AME2012LD8. It should be noted that there are many zero and small terms of $\sigma_{a_{i}}B_{i,Z,A}$ in the Eq.~(\ref{LD8stat}). The simulated distribution has rather high peak at the center which is not expected in a  realistic situation. If zero and small $\sigma=\sigma_{a_{i}}B_{i,Z,A}$ terms in Eq.~(\ref{LD8stat}) is neglected, the simulated distribution is quite similar to the normal statistical distribution. The normal assumption of the statistical distribution is a kind of renormalization of the simulated distribution.

More tests can be performed to see that the normal assumption of the statistical and systematic distribution is acceptable. The liquid drop model is more suitable to describe the heavy nuclei than the light nuclei. One can assume that the residues from the heavy nuclei of the LD8 is more contributed by the statistical uncertainty while that from the light nuclei is more contributed by the systematic uncertainty. Table~\ref{normaltest} presents two widely used normal test methods, Lilliefors~\cite{Lilliefors1967} and Jarque-Bera tests~\cite{Jarque1980}, on the residues of the light ($8\leq Z\leq 20$), heavy ($Z\geq 90$), and all nuclei in  AME2012LD8. The residues of all nuclei are rejected to be taken from a normal distribution at $1\%$ significant level (the significant level for rejection is smaller than $0.1\%$). The residues of the light and heavy nuclei can not be rejected at $1\%$ and $5\%$ (except the Jarque-Bera test for heavy nuclei) significant levels. The normal distribution is an acceptable estimation (not exact) for the distributions of the residues of light and heavy nuclei, and the distributions of systematic and statistical uncertainties.

\begin{table}
\caption{\label{normaltest} The Lilliefors and Jarque-Bera normal tests for the residues of the binding energies of the light, heavy, and all nuclei in AME2012LD8.}
\begin{ruledtabular}
\begin{tabular}{ccc}
 data  & test method          & \emph{p}-value  \\ \hline
 all   & Lilliefors test      & $\leq$0.1\%                \\
 all   & Jarque-Bera test     & $\leq$0.1\%                \\
 light & Lilliefors test      & 28.4\%                \\
 light & Jarque-Bera test     & 10.7\%                \\
 heavy & Lilliefors test      & 8.4\%                 \\
 heavy & Jarque-Bera test     & 2.0\%                 \\
\end{tabular}
\end{ruledtabular}
\end{table}

\begin{figure}
\includegraphics[scale=0.30]{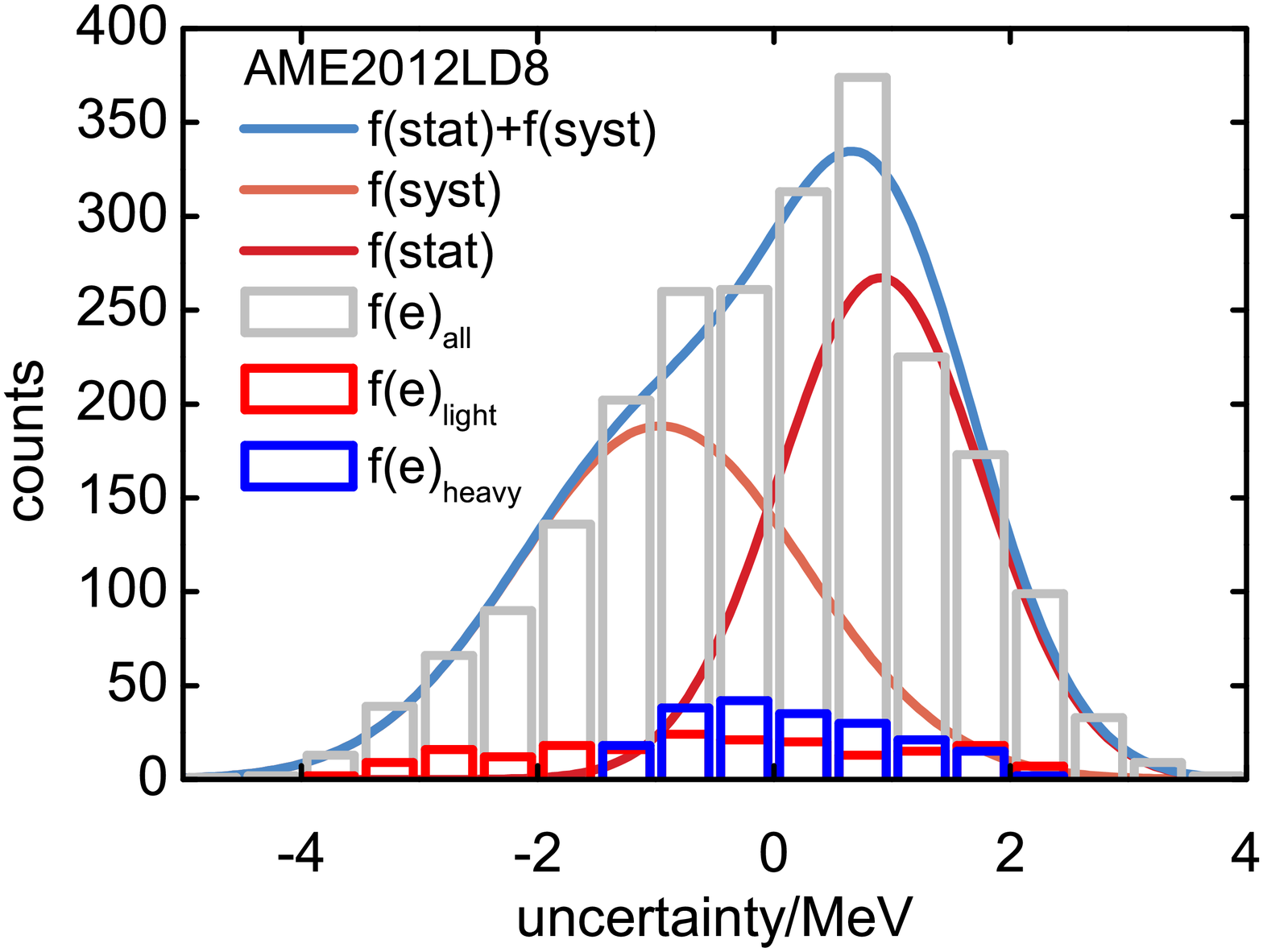}
\includegraphics[scale=0.30]{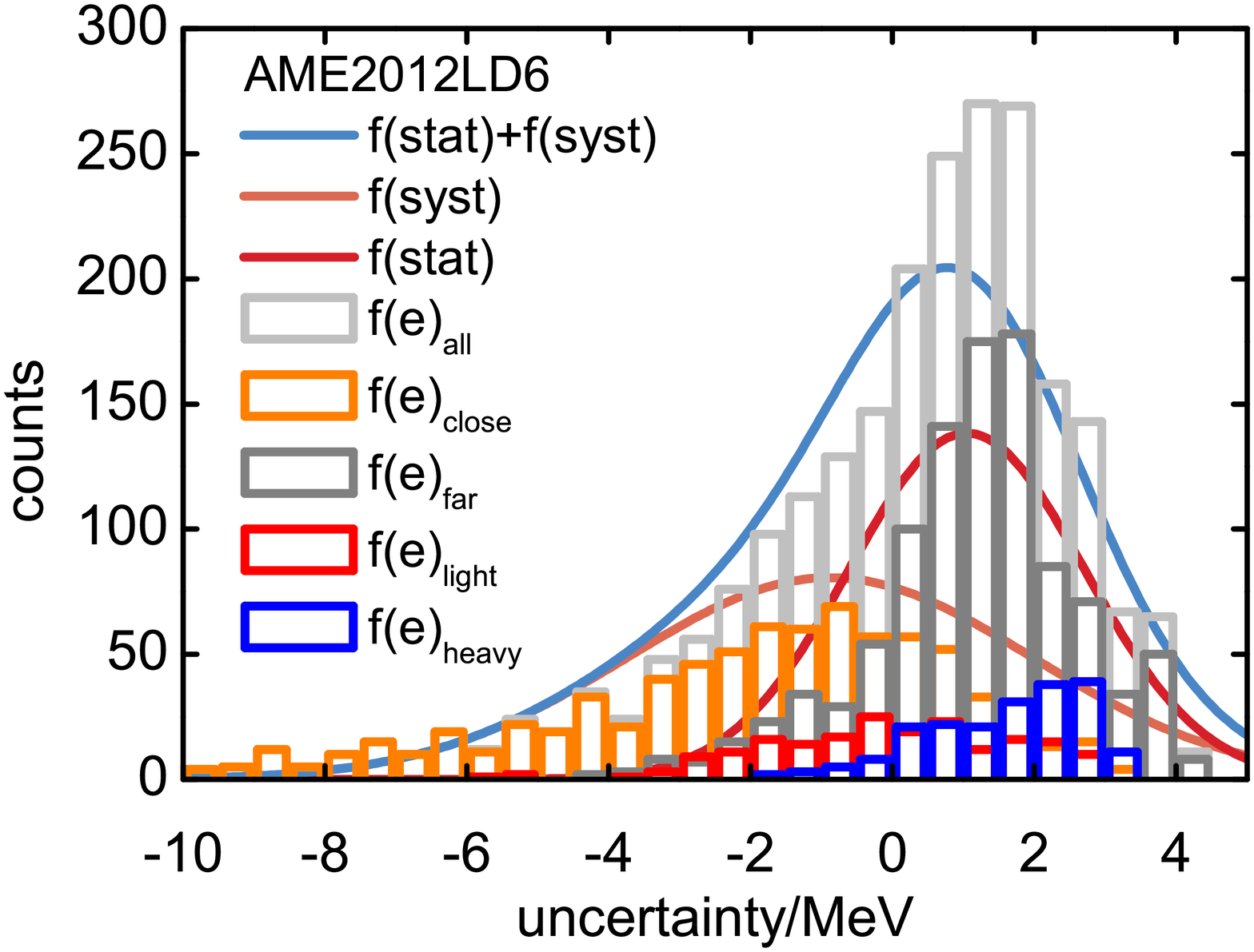}
\caption{\label{eachdistribution}(Color online) The distribution of the residues of the binding energies of the light and heavy nuclei, nuclei close to and far from shell (only for AME2012LD6), all nuclei, and corresponding estimated uncertainties of AME2012LD8 and AME2012LD6.}
\end{figure}

The validity of the relationship of the Eq.~(\ref{residue}) can be further investigated through the consideration of the specific nuclei. Besides the light and heavy nuclei, one can assume that the residues of the nuclei close to and far from shell are expected to be more contributed by the systematic and statistical uncertainties in LD6, respectively. Figure~\ref{eachdistribution} presents the distribution of the residues of the light and heavy nuclei in AME2012LD8 and AME2012LD6, and those of the nuclei close to and far from shell in AME2012LD6. The ``close to'' and ``far from'' shell are defined by the value of $\Sigma_{Z_{0},N_{0}}e^{-\frac{(Z-Z_{0})^{2}+(N-N_{0})^{2}}{a_{0}^{2}}}$ in the Eq.~(\ref{shell2}), larger than $0.1$ and smaller than $0.01$, respectively. It is clearly seen that the distributions of these specific nuclei do agree with the physical consideration, the residues of light and close to shell nuclei, $f(e)_{light}$ and $f(e)_{close}$, are mostly distributed inside the $f(syst)$, and those of heavy and far from shell nuclei, $f(e)_{heavy}$ and $f(e)_{far}$, are mostly inside the $f(stat)$. Although $f(e)_{heavy}$ in the LD8 and $f(e)_{light}$ in the LD6 distribute in the mixed region, their centroid are more closed to the centroid of $f(stat)$ and $f(syst)$, respectively. Figure~\ref{eachdistribution} provides an examination on the Eq.~(\ref{residue}). The shell effect is the most important systematic uncertainty in LD6, which strongly underestimate the binding energies of the nuclei close to shell. Almost all large negative residues of LD6 come from the shell effect. The relationship $f(e)=f(e)_{close}=f(syst)$ is well satisfied at the region smaller than $-4$ MeV, where the $f(stat)$ becomes zero. At another side, an approximately $f(e)=f(e)_{far}=f(stat)$ can be found at the region larger than $2$ MeV, of which the deviation comes from the tail of $f(syst)$.

\begin{table*}
\caption{\label{noshellheavysd} Four estimated normal parameters, the variance of the estimated uncertainties, the mean value and standard deviations of the residues of the binding energies of the light and heavy nuclei, the nuclei close to and far from shell in AME2012LD6. All data are in the unit of MeV, except the $\frac{1}{2}(m_{stat}^{2}+\sigma_{stat}^{2})$ and $\frac{1}{2}(m_{syst}^{2}+\sigma_{syst}^{2})$ in MeV$^{2}$.}
\begin{ruledtabular}
\begin{tabular}{cccccccccc}
 data  & $m_{stat}$  & $\sigma_{stat}$  &$m_{syst}$  & $\sigma_{syst}$ & $\frac{1}{2}((m_{stat}-m)^{2}+\sigma_{stat}^{2})$ &$\frac{1}{2}((m_{syst}-m)^{2}+\sigma_{syst}^{2})$ &$m$& $\sigma$\\
\hline
close to shell & 0.06    &  0.87   &  -4.00   &  2.66 &  2.44 & 5.59  & -1.97 & 2.83 \\
light          & 1.41    &  0.47   &   -1.59  &  1.08 &  1.23 & 1.71  & -0.08 & 1.72 \\
far from shell &  0.54   &  1.83   &   1.71   & 0.70  & 1.84  &  0.42 & 1.12  & 1.50 \\
heavy          &         &  2.62   &          &       &       &       & 1.60  & 1.13 \\
\end{tabular}
\end{ruledtabular}
\end{table*}

The fitting procedure finds a balance for the statistical and systematic uncertainty. The normalized factor in Eq.~(\ref{residue}) is expected to be $\frac{1}{2}$ in the best fitted case. The residues of the specific nuclei may be more contributed by one kind of the uncertainty. It is interesting to see the agreement between the above discussions and the UDM results from these specific regions with the same normalized factor $\frac{1}{2}$. Table~\ref{noshellheavysd} presents the UDM results, mean value $m$, and the standard deviation $\sigma$ of the four specific regions in AME2012LD6. It should be noted that the centroid of the residues for the specific regions may deviate largely from zero. The variances of the statistical and systematic uncertainties are $\frac{1}{2}((m_{stat}-m)^{2}+\sigma_{stat}^{2})$ and $\frac{1}{2}((m_{syst}-m)^{2}+\sigma_{syst}^{2})$. It is seen that the contribution of the systematic and statistical uncertainties are large for the residues of nuclei close to and far from shell, respectively. If the normalized factors of corresponding uncertainties are increased, the value of the criteria of the UDM become more close to zero and the contribution of the uncertainty become larger, which are expected.

It should be emphasized that the statistical variance simulated through the Eq.~(\ref{LD8stat}) is meaningful in the global region, not in the specific regions, such as the heavy and light nuclei. Because the parameters of the LD and their standard deviations are fitted to all nuclei, a representative sample including all degrees of freedom of the model, which mean $A$, $A^{2/3}$, $Z(Z-1)A^{-1/3}$, $I(I+1)/A$, $I(I+1)/A^{4/3}$, and $\delta A^{-1/2}$ in the Eq.~(\ref{LD6}). They are expected to be suitable for representative nuclei including all of these degrees of freedom, such as nuclei close to and far from shell distributing in all region on the chart of nuclide. They may not be suitable for specific region on the chart of nuclide, such as the heavy and light nuclei. The value of $\sigma_{stat}$ of the nuclei far from shell agrees well with the $\sigma$ even if systematic uncertainty is not considered. The $\sigma_{stat}$ may be overestimated and underestimated for the heavy and light nuclei, respectively, seen from the Table~\ref{noshellheavysd}. The estimated statistical variance of the residues of the heavy nuclei $\sigma_{stat}^{2}>2\sigma^{2}$ is beyond the ability of the UDM. One possible explanation is that one can not clarify the systematic uncertainty if the statistical uncertainty is very large. It can be seen from Eq.~(\ref{LD8stat}) that the simulated statistical variance become larger as A increases because of increasing $\sigma_{a_{v}}A$ and $\sigma_{a_{s}}A^{2/3}$, and smaller when A decreases. The heavy and light nuclei are not representative samples because they only include large or small $A$ when the parameters are obtained from all possible $A$. In Ref.~\cite{erler2011}, the statistical uncertainty of the two-neutron drip line estimated from energy density functionals increases from the light to heavy nuclei.

The centroid of $f(stat)$ of AME2012LD6 is estimated to be $1.05$ MeV, seen from Table~\ref{msd}, deviates from the estimated $m_{stat}$ in the Table~\ref{noshellheavysd} ($m_{stat}$ is estimated as $m$ for the heavy nuclei). It indicates that the normal assumption of $f(stat)$ is an estimation of the summation of many distributions with different centroids and standard deviations.

\section{\label{sec:level5}Investigations on unmeasured nuclei}

\begin{figure}
\includegraphics[scale=0.30]{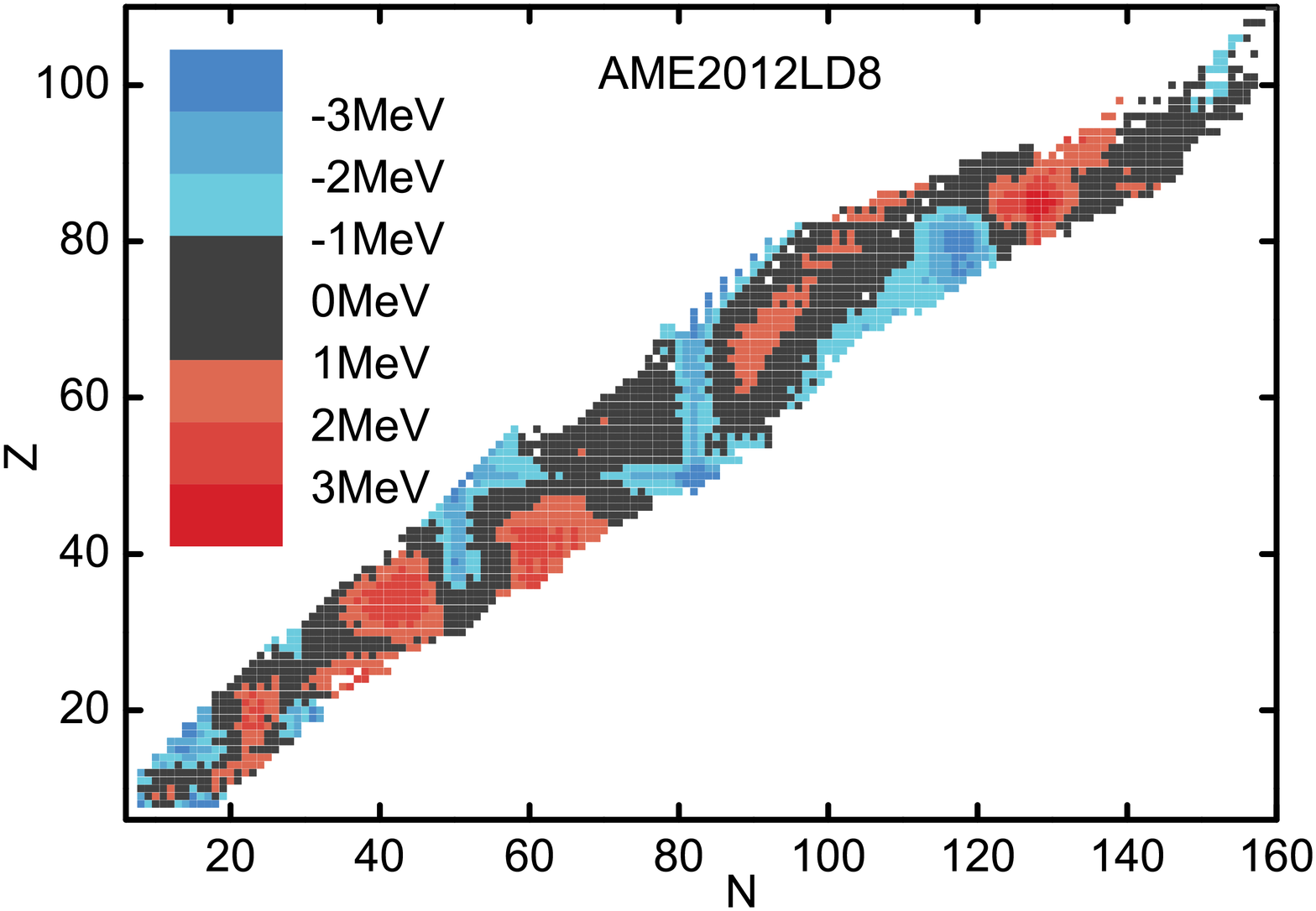}
\includegraphics[scale=0.30]{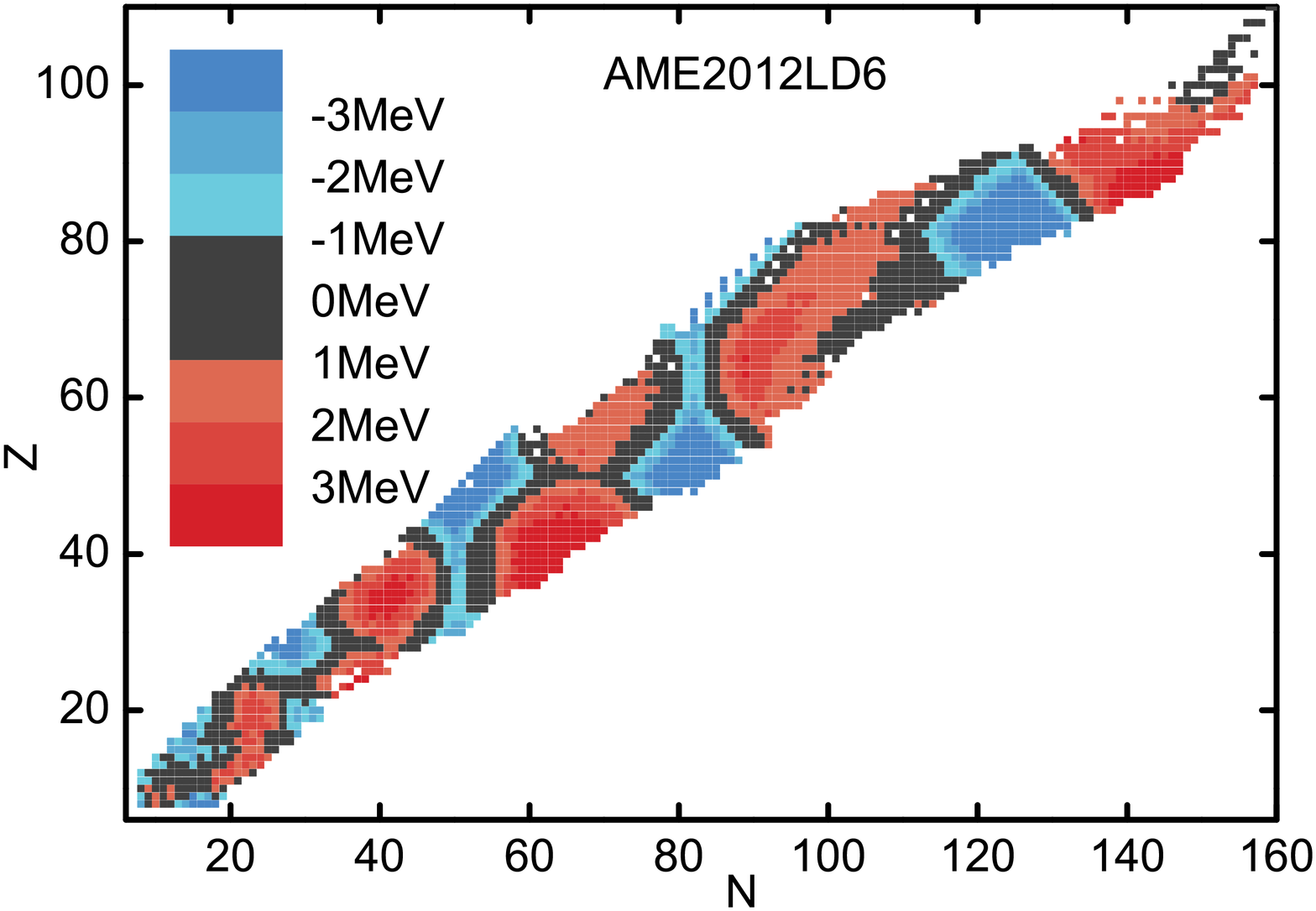}
\caption{\label{LDSD} (Color online) The distribution of the residues of the binding energies on the chart of nuclide of the AME2012LD8 and AME2012LD6.}
\end{figure}

Figure~\ref{LDSD} presents the residues of AME2012LD6 and AME2012LD8. It is clearly seen that the residues between the LD and the observed data is rather locally dependent, such as the nuclei close to shell have larger negative residues and heavy nuclei have positive residues in the LD6. The simulated systematic uncertainty may be not suitable for the specific regions of nuclei as discussed before. It is interesting to further investigate on how present method can be applied to unmeasured nuclei, such as the extreme neutron-rich nuclei. Table~\ref{neutronrich} shows the estimated statistical uncertainties for the nuclei from $(A_{max}+1,Z)$ to $(A_{max}+5,Z)$ in AME2012LD8 and AME2012LD6, where $(A_{max},Z)$ is the heaviest nuclei for each isotopes. The mean values $m_{stat}$ in Table~\ref{neutronrich} are taken to be the same as estimated in Table~\ref{msd}. The estimated statistical uncertainties are slightly larger than those of the measured nuclei because of larger $A$. It is acceptable to use such statistical uncertainties for the global investigation of the unmeasured neutron-rich nuclei, but with relatively large uncertainties. Better choices are discussed for specific cases.

\begin{table}
\caption{\label{neutronrich} The mean values (the same as Table~\ref{msd}), the estimated standard deviations, and the variance of the statistical uncertainties nuclei for the nuclei from $(A_{max}+1,Z)$ to $(A_{max}+5,Z)$ in AME2012LD8 and AME2012LD6, where $(A_{max},Z)$ is the heaviest nuclei for each isotopes. All data are in the unit of MeV, except the $\frac{1}{2}(m_{stat}^{2}+\sigma_{stat}^{2})$ in MeV$^{2}$.}
\begin{ruledtabular}
\begin{tabular}{cccc}
 model           &$m_{stat}$  & $\sigma_{stat}$  & $\frac{1}{2}(m_{stat}^{2}+\sigma_{stat}^{2})$   \\ \hline
 LD8 & 0.91	& 1.01	& 0.92  \\
 LD6 & 1.05 & 1.95	& 2.46  \\
\end{tabular}
\end{ruledtabular}
\end{table}

One possible choice is to calculate the mean values and the standard deviations for the measured nuclei around the unmeasured ones. Then use these two values to modified the calculations. Such as the mean value and the standard deviation of the residues are $1.60$ and $1.13$ MeV for heavy nuclei in AME2012LD6, respectively. The calculated $BE(A,Z)_{LD6}$ of an unmeasured nucleus in this region is expected to be modified as $BE(A,Z)_{LD6}-1.60\pm1.13$ MeV, which has much less standard deviation compared with that of the all residues, $2.52$ MeV. As seen in FIG.~\ref{eachdistribution}, the distributions of specific nuclei are generally much smaller than the distribution of the residues except the nuclei close to shell.

\begin{figure}
\includegraphics[scale=0.30]{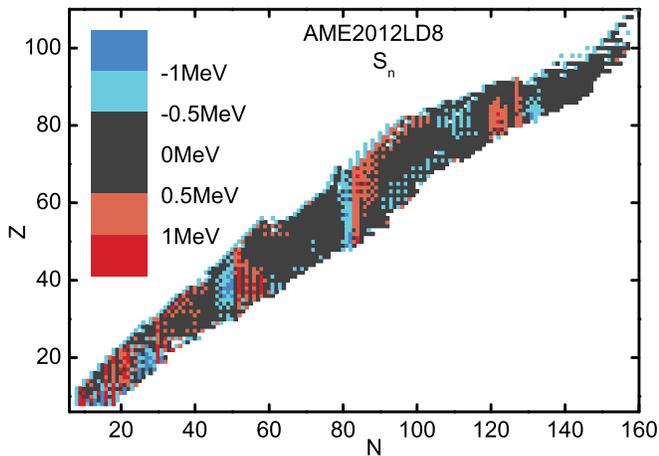}
\caption{\label{LDSnSD} (Color online) The distribution of the residues of the neutron separation energies on the chart of nuclide of the AME2012LD8. }
\end{figure}

\begin{table*}
\caption{\label{LDSn} Four estimated normal parameters, the variance of the estimated uncertainties, the mean value and standard deviation of the residues of the neutron separation energies in AME2012LD8 and AME2012LD6. All data are in the unit of MeV, except the $\frac{1}{2}(m_{stat}^{2}+\sigma_{stat}^{2})$ and $\frac{1}{2}(m_{syst}^{2}+\sigma_{syst}^{2})$ in MeV$^{2}$.}
\begin{ruledtabular}
\begin{tabular}{cccccccccc}
 Model &data  &$m_{stat}$  & $\sigma_{stat}$  &$m_{syst}$  & $\sigma_{syst}$ & $\frac{1}{2}(m_{stat}^{2}+\sigma_{stat}^{2})$ &$\frac{1}{2}(m_{syst}^{2}+\sigma_{syst}^{2})$ &$m$& $\sigma$\\
\hline
 LD8 &all   & 0.06    & 0.67    &   -0.01  &  0.31 &  0.23 &  0.05 & 0.03 & 0.52 \\
 LD8 &heavy &         & 1.06    &          &       &       &       & -0.03& 0.27 \\
 LD8 &light & 1.05    & 0.16    &  -0.83   &  0.41 &  0.67 & 0.72  & 0.11 & 1.00 \\
 LD6 &all   &         & 1.31    &          &       &       &       & 0.07 & 0.66 \\
\end{tabular}
\end{ruledtabular}
\end{table*}

Another choice is to investigate the neutron separation energies, which have smaller residues and less local dependance than those of binding energies, seen from FIG.~\ref{LDSnSD}. $S_{n}(A,Z)=BE(A,Z)-BE(A-1,Z)$ reduces the systematic uncertainties because many neighbour nuclei have similar systematic trends, such as the shape and shell effect. The UDM results of the neutron separation energies of AME2012LD8 and AME2012LD6 in Table~\ref{LDSn} show that the residues are mostly contributed by the statistical uncertainties. $\sigma_{stat}$ of the heavy and light nuclei have similar situation to the Table~\ref{noshellheavysd}, larger and smaller than that of all nuclei, respectively. But the increment of $\sigma_{stat}$ from all nuclei to heavy nuclei of the neutron separation energies (58\%) is much smaller than that of the binding energies (200\%). It is a better choice to discuss the uncertainty of the separation energies rather than that of the binding energies.

The unmeasured super heavy nuclei (SHE) are very important. The present method has its limitation on the discussion on the SHE. The LD give nice description on the heavy nuclei, small $\sigma$ but much larger estimated uncertainty $\sigma_{stat}$, seen from Table~\ref{noshellheavysd} and \ref{LDSn}. Better methods to estimate the uncertainty of heavy nuclei should be considered in the future. The further discussion on the drip line and the SHE are not scheduled in the present work.

\section{\label{sec:level6}Start from stability line}

\begin{table*}
\caption{\label{parastable} Four sets of parameters including standard deviation of each parameter in parentheses for the LD5, LD6, LD7, and LD8 fitted to the nuclei around stability line in AME2012, the mean value $m(e)$, standard deviation $\sigma(e)$, skewness to the power of $1/3$, $p_{3}^{1/3}(e)$, and kurtosis to the power of $1/4$, $p_{4}(e)$, of the residues of the binding energies of the nuclei around stability line and all measured nuclei. All data are in the unit of MeV except the dimensionless quantity $a_{0}^{2}$.}
\begin{ruledtabular}
\begin{tabular}{cccccccccccccc}
 Model &data  &$a_{v}$  & $a_{s}$  & $a_{c}$  & $a_{av}$ & $a_{as}$  & $a_{p}$ & $a_{sh}$ & $a_{0}^{2}$ & $m(e)$ & $\sigma(e)$  &$p_{3}^{1/3}(e)$ &$p_{4}^{1/4}(e)$\\
\hline
  LD5 &stability & 16.231(50)  & 19.17(16)     &   0.7590(38)  &  23.25(15)&            & 12.1(1.5) &          &      &-0.02& 2.59 & -2.63 & 3.81  \\
  LD5 &all&      &      &      &    &     &   &  &                                                                    &-1.78& 4.17 & -5.80  & 7.79  \\
  LD6 &stability & 15.786(90) & 18.13(23)     &   0.7151(83)  &  28.11(84)& 32.9(5.6)  & 11.8(1.4) &          &       &-0.01& 2.52 & -2.71 & 3.76  \\
  LD6 &all&      &      &      &    &     &   &  &                                                                    &-0.32& 2.67 & -2.73  & 3.83  \\
  LD7 &stability & 16.282(30) & 19.387(95)    &   0.7590(23)  & 23.757(91)&          & 11.96 (88) &    0.0536(17) &40 &-0.02& 1.57 & -1.05  & 2.03  \\
  LD7 &all&      &      &      &    &     &   &  &                                                                    &-1.74& 3.74 & -6.01  & 8.28  \\
  LD8 &stability & 15.671(48) & 17.97(12)     &   0.6987(44)  &  30.46(45)& 45.3(3.0)  & 11.56(74) &0.0559(14) &40    &-0.01& 1.33 & -0.88  & 1.69  \\
  LD8 &all&      &      &      &    &     &   &  &                                                                    &0.26 & 1.49 & 0.65  & 1.92  \\
\end{tabular}
\end{ruledtabular}
\end{table*}

\begin{table*}
\caption{\label{UDMstable} Four estimated normal parameters, the variance of the estimated uncertainties, the mean value and standard deviations of the residues of the binding energies of nuclei around stability line and all measured nuclei with four sets of the parameters obtained from LD5, LD6, LD7, and LD8 fitted to the nuclei around stability line in AME2012. All data are in the unit of MeV, except the $\frac{1}{2}(m_{stat}^{2}+\sigma_{stat}^{2})$ and $\frac{1}{2}(m_{syst}^{2}+\sigma_{syst}^{2})$ in MeV$^{2}$.}
\begin{ruledtabular}
\begin{tabular}{cccccccccc}
 Model &data  &$m_{stat}$  & $\sigma_{stat}$  &$m_{syst}$  & $\sigma_{syst}$ & $\frac{1}{2}((m_{stat}-m)^{2}+\sigma_{stat}^{2})$ &$\frac{1}{2}((m_{syst}-m)^{2}+\sigma_{syst}^{2})$ &$m$& $\sigma$ \\
\hline
 LD5 &stability &          & 4.00    &          &       &       &      &-0.02& 2.59 \\
 LD5 &all       & -1.47    & 4.13    & -2.09    &  4.20 &  8.59 &  8.81&-1.78& 4.17\\
 LD6 &stability &          & 5.07    &          &       &       &      &-0.01& 2.52 \\
 LD6 &all       &          & 5.22    &          &       &       &      &-0.32& 2.67 \\
 LD7 &stability &          & 2.22    &          &       &       &      &-0.02& 1.57 \\
 LD7 &all       & -0.30    & 2.29    & -3.19    &4.30   &3.68   &10.30  &-1.74& 3.74\\
 LD8 &stability &          & 2.49    &          &       &       &      &-0.01& 1.33 \\
 LD8 &all       &          & 2.57    &          &       &       &      &0.26 & 1.49 \\
\end{tabular}
\end{ruledtabular}
\end{table*}

\begin{figure}
\includegraphics[scale=0.30]{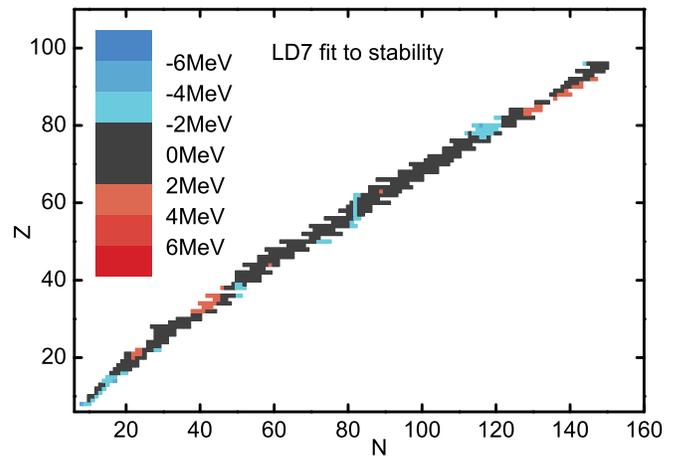}
\includegraphics[scale=0.30]{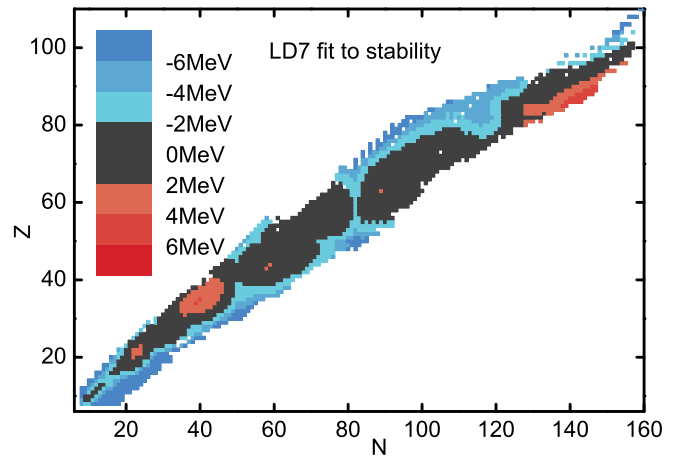}
\caption{\label{LD7sd} (Color online) The distribution of the residues of the binding energies on the chart of nuclide of the LD7 fitted to the nuclei around the stability line in AME2012. Only the nuclei around stability line are shown in the upper panel, while the others are together shown in the lower panel. }
\end{figure}

\begin{figure}
\includegraphics[scale=0.30]{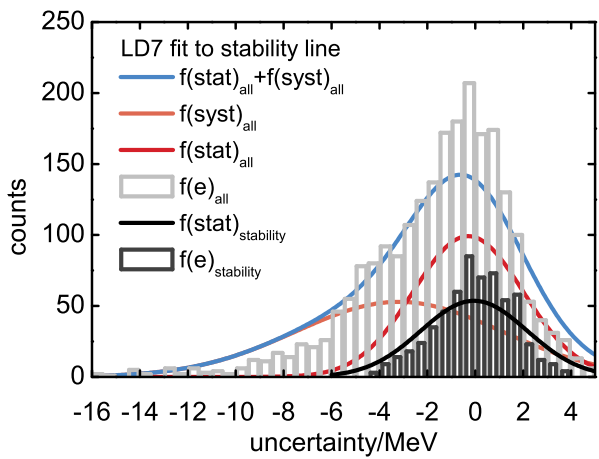}
\caption{\label{LD7counts} (Color online) The distribution of the residues of the binding energies of the nuclei around stability line and all nuclei, and the corresponding estimated uncertainties of the LD7 fitted to the nuclei around stability in AME2012.}
\end{figure}
 	 	 	
It is interesting to see what the LD and UDM results are if only the nuclei around stability line are known, which is helpful for the understanding how the unmeasured nuclei can be described through the present observed data. $597$ nuclei are selected from AME2012, which are around the stability line $Z=\frac{A/2}{1+0.0077A^{2/3}}$~\cite{heydebasic} and with the small observed uncertainties of the binding energies (in general, $\triangle BE_{expt}<1$ keV at $8\leq Z \leq40$, $<2$ keV at $41\leq Z \leq70$, and $<3$ keV at $Z \geq71$). Table~\ref{parastable} presents the parameters obtained from these data for LD8, LD6, LD7, and LD5. The fittings of the LD8 and LD6 give large uncertainty on the surface asymmetry term. The LD7 and LD5 are then fitted without this term, defined as follow, respectively: 
\begin{eqnarray}\label{LD7}
BE(A,Z)_{LD7} &=& a_{v}A-a_{s}A^{2/3}-a_{c}Z(Z-1)A^{-1/3} \nonumber \\
    & & -a_{a}^{v}I(I+1)/A+BE(Z,A-Z)_{shell} \nonumber \\
    & & +\delta a_{p}A^{-1/2},
\end{eqnarray}
\begin{eqnarray}\label{LD5}
 BE(A,Z)_{LD5} &=& a_{v}A-a_{s}A^{2/3}-a_{c}Z(Z-1)A^{-1/3} \nonumber \\
    & & -a_{a}^{v}I(I+1)/A+\delta a_{p}A^{-1/2}.
\end{eqnarray}

It is obvious that the nuclei around stability line are less representative than all measured nuclei with larger uncertainties in each parameters of the LD, partially because of the less number of the data and partially because of the different function of each parameter. For example, both the paring and shell terms are less correlated to the isospin and other terms. The uncertainty of these two parameters changes not much for the data from the stability line to all measured nuclei because the most important change is on the isospin degree of freedom. The uncertainty of the volume and surface asymmetry term changes a lot as the isospin changes. The uncertainty of the surface asymmetry term is much larger than that of the volume asymmetry term because $I(I+1)/A^{4/3}$ is generally much smaller than $I(I+1)/A$. Thus the uncertainty of the surface asymmetry term is very large (around $10\%$) if only nuclei near stability line is considered. It is stated in Ref.~\cite{myers1966} that the form of the surface symmetry term is obtained not from any evidence of the nuclear mass but from the form of the volume asymmetry term.

The UDM results of these sets of parameters of the LD are presented in the Table~\ref{UDMstable}. Because of the large uncertainties of the parameters, the estimated $\sigma_{stat}^{2}$ is larger than $2\sigma^{2}$ in most cases, which is the limitation of the UDM. Although the systematic uncertainties can not be determined among the large statistical uncertainties, many interesting discussions can be addressed.

The standard deviations $\sigma$ of the nuclei around stability line decrease a little from the LD5 to LD6 and from the LD7 to LD8 because of the added surface asymmetry term while the statistical uncertainty increases. It indicates that this term is not very necessary for the nuclei around stability line, which agrees with the original form of the LD. In the LD6 and LD8, $\sigma$ of the residues from stability line are similar compared with to that from all measured nuclei. Although the surface asymmetry term can be excluded when the data is around stability line and its uncertainty is rather large if included, its value is acceptable obtained (the best fitted values in Table~\ref{para} are inside $a_{as}\pm2\sigma_{a_{as}}$ in Table~\ref{parastable}) and very useful for the prediction. It is reasonable because the surface asymmetry term has its physical meaning on the isospin degree of freedom. The real uncertainty is less than the estimation. The situation of the heavy nuclei is similar. Although the estimated uncertainty is large, the real uncertainty is much smaller for both the binding energies and the neutron separation energies because the liquid drop assumption is suitable for the heavy nuclei.

In the LD5 and LD7, $\sigma$ of the residues from all measured nuclei are dramatically larger compared with that from stability line. It is expected because the nuclei near stability line, $Z=\frac{A/2}{1+0.0077A^{2/3}}$, is nearly one dimension on the chart of nuclei, which misses many terms on isospin degree of freedom correlated to other degrees of freedom. Such as, for certain $Z$, $A-2Z$ is limited around $\frac{0.0077A^{2/3}}{1+0.0077A^{2/3}}$. The correlation between $Z$ and other $A-2Z$ are missing. Figure~\ref{LD7sd} presents the comparison of the results of the LD7 for both the nuclei around stability line and all measured nuclei. It is clearly seen that the parameters obtained from the stability line fail when approaching to neutron- and proton-rich nuclei. It is nice to see that the standard deviations of the statistical uncertainties $\sigma_{stat}$ keep almost the same from the stability line to all measured nuclei in both the LD5 and LD7. Such estimated $\sigma_{stat}$ can be used in the UDM and give nice description on the residues, seen from FIG.~\ref{LD7counts} taking the LD7 for example. Although estimated $\sigma_{stat}$ seem to be large when only data around stability line is concentrated, their values are indeed useful for predictions in the global  chart of nuclide even if some important terms are missing, such as the surface asymmetry term. But of course the systematic uncertainty may be very large. It is expected that the estimated $\sigma_{stat}$ obtained from the present observed data in Sec.~\ref{sec:level3} is useful to scale the statistical uncertainty for unmeasured nuclei. But as discussed before, such estimations are more suitable for a global investigation, may not for local cases.

In FIG.~\ref{LD7counts}, the normalized factor of $f(stat)_{stability}$ is set to be one, which shows that the statistical uncertainty is very large that the systematic uncertainty can not be clarified for the nuclei around the stability line. It is seen that the systematic uncertainty less contributes to the residues of these nuclei with the same set of parameters when the data changes to all measured nuclei.

\section{\label{sec:level7}Summary}

In conclusion, a method is suggested to decompose the statistical and systematic uncertainties of a theoretical model after fitting procedure. The two uncertainties are obtained through the total uncertainty and the model parameters obtained from the fitting. Such uncertainty decomposition method (UDM) are applied to the liquid drop model (LD) as an example. The estimated distribution can well reproduce the distribution of the residues of the binding energies. The specific nuclei locate where the physical considerations expect in the estimated distribution obtained without these considerations. Such as the light nuclei and nuclei close to shell locate mostly inside the distribution of the systematic uncertainty, while the heavy nuclei and nuclei far from shell inside the distribution of the statistical uncertainty. The results are obtained purely from the mathematic forms of the LD and the observed data. Thus the UDM may be useful for the discovery of hinted physics in certain cases. The validity of the UDM is tested through various approaches. It is acceptable that the distribution of the statistical and systematic uncertainties are assumed to be normal. 

The present work is constrained in the simple model under the simple statistical assumptions to see what can be obtain from the residues. It should be noted that the realistic distributions of the statistical uncertainty from the model parameters and the parameters themselves are still not clearly known for the present simple model and some modern nuclear mass model, such as the Hartree-Fock-Bogoliubov model~\cite{Goriely2009,Goriely2013} and the Weizs\"{a}ecker-Skyrme mass model~\cite{liu2011,Wang2014}. Some advanced method to deal with the systematic uncertainty of the mass model are of great interesting, such as the image reconstruction techniques~\cite{Morales2010}, the radial basis function approach~\cite{Wang2011}, and the method for the evaluation of AME2012~\cite{audi2012,audi20121}. The former two reconstruct the uncertainty through the systematic trends of the residues, while the latter evaluates huge amount of the observed values and their uncertainties.

The present work also shows the effectiveness and limitation of the UDM in the investigation of the unmeasured nuclei. In a global view, the statistical uncertainty are well estimated even when certain important terms in the model are missing. The statistical uncertainty may be overestimated or underestimated in the specific regions of nuclei, such as the heavy and light nuclei. More statistical methods combined with the physical considerations should be considered to simulate the uncertainties of the parameters and the statistical uncertainties.

The present work shows that the residues indeed includes more information beyond the most used two, the mean value and the standard deviation, which are rarely discussed before. It is expected that the UDM can be used in other theoretical works, such as fitting effective nuclear force to nuclear data, which is helpful for the study of the statistical and systematic uncertainties of the levels through the nuclear shell model.

\section{\label{sec:level8}Acknowledgement}

The author acknowledge to the useful suggestion from Dr. Chong Qi,
and the collection of the data by the
students enrolled the course ``applied statistics'' in 2013. This
work has been supported by the National Natural Science Foundation
of China under Grant No.~11305272, the Specialized Research Fund for the Doctoral Program
of Higher Education under Grant No.~20130171120014, the Fundamental Research Funds
for the Central Universities under Grant No.~14lgpy29, the Guangdong Natural Science Foundation under Grant No.~2014A030313217, and the Pearl River S\&T Nova Program of Guangzhou under Grant No.~201506010060.

\end{document}